\documentclass[aps,prb,twocolumn,amsmath,amssymb,groupedaddress,floatfix]{revtex4}
\usepackage[utf8]{inputenc}
\usepackage{xcolor}
\usepackage{amssymb}
\usepackage{amsmath}
\usepackage{bbold}
\usepackage{graphicx}

\newcommand{\lamc}{{\Lambda_{\rm c}}}
\newcommand{\eps}{\epsilon}
\newcommand{\up}{\uparrow}
\newcommand{\down}{\downarrow}
\newcommand{\bs}[1]{\mathbf{#1}}

\newcommand{\bk}{\mathbf{k}}
\newcommand{\bq}{\mathbf{q}}
\newcommand{\bQ}{\mathbf{Q}}
\begin{document}
\title{Dynamical functional renormalization group computation of order parameters \\
 and critical temperatures in the two-dimensional Hubbard model}
%
\author{Demetrio Vilardi}
\author{Pietro M. Bonetti}
\author{Walter Metzner}
\affiliation{Max Planck Institute for Solid State Research, Heisenbergstrasse 1, D-70569 Stuttgart, Germany}
\date{\today}
\begin{abstract}
We analyze the interplay of antiferromagnetism and pairing in the two dimensional Hubbard model with a moderate repulsive interaction. Coupled charge, magnetic and pairing fluctuations above the energy scale of spontaneous symmetry breaking are treated by a functional renormalization group flow, while the formation of gaps and order below that scale is treated in mean-field theory. The full frequency dependences of interaction vertices and gap functions is taken into account.
We compute the magnetic and pairing gap functions as a function of doping $p$ and compare with results from a static approximation. In spite of strong frequency dependences of the effective interactions and of the pairing gap, important physical results from previous static functional renormalization group calculations are confirmed. In particular, there is a sizable doping regime with robust pairing coexisting with N\'eel or incommensurate antiferromagnetism. The critical temperature for magnetic order is interpreted as pseudogap crossover temperature.
Computing the Kosterlitz-Thouless temperature from the superfluid phase stiffness, we obtain a superconducting dome in the $(p,T)$ phase diagram centered around 15 percent hole doping.
\end{abstract}

\maketitle

%
\section{Introduction}
Shortly after the discovery of high-temperature superconductivity in cuprates, Anderson \cite{Anderson1987} proposed the two-dimensional Hubbard model to describe the behavior of the valence electrons in the copper-oxygen planes. Indeed, the model captures the most prominent ordered phases observed in high-$T_c$ cuprates: antiferromagnetism and $d$-wave superconductivity. \cite{Scalapino2012}

Antiferromagnetism in the Hubbard model is not always of N\'eel type, that is, with antiparallel spin orientation between adjacent lattice sites. Magnetic order with (generally incommensurate) wave vectors away from the N\'eel point $(\pi,\pi)$ has been found away from half-filling in several mean-field studies, \cite{Machida1989,Schulz1990,Dombre1990,Fresard1991,Igoshev2010} and also, including fluctuations, by expansions in the hole-density. \cite{Shraiman1989,Chubukov1992,Chubukov1995,Kotov2004}

Unbiased evidence for superconductivity with a sizable energy scale already at moderate interaction strengths has been established from functional renormalization group (fRG) calculations, \cite{Zanchi2000,Halboth2000,Honerkamp2001,Honerkamp2001a,Metzner2012,Eberlein2014} and from quantum cluster methods at intermediate to strong coupling. \cite{Lichtenstein2000, Maier2000,Maier2005,Aichhorn2006,Capone2006,Kancharla2008,Gull2013,Zheng2016}
Recently, a fRG flow starting from the dynamical mean-field solution (instead of the bare action) has confirmed robust pairing with $d$-wave symmetry at strong coupling. \cite{Vilardi2019}

The fRG flow is defined by a successive, scale-dependent integration of the fermionic fields in a path integral representation of the effective action. \cite{Metzner2012} Spontaneous symmetry breaking is signaled by a divergence of effective interactions at a critical energy scale $\Lambda_c$. Accessing the ordered phase by continuing the flow beyond the critical scale is possible, \cite{Salmhofer2004} but rather complicated due to a proliferation of anomalous interaction terms. \cite{Gersch2008,Eberlein2013}
One option is to decouple the fermionic interaction by introducing a bosonic order parameter field via a Hubbard-Stratonovich transformation, and to study a coupled flow involving fermions and bosons. \cite{Baier2004}
Alternatively, as a ``poor man's'' approximation, one may restrict the effective interactions below the critical scale $\Lambda_c$ to those (reduced) interactions which drive the symmetry breaking. \cite{Reiss2007,Wang2014} The flow of these interactions is governed only by a single channel (one for each order parameter), and the final order parameters at the end of the flow are given by simple gap equations, which can be derived from the fermionic flow equation \cite{Wang2014} or by introducing a bosonic order parameter field at the critical scale. \cite{Bonetti2020} Consistently formulated, this procedure is equivalent to a mean-field theory (MFT) for the degrees of freedom below the critical scale. In a regime where pairing is the only instability, it turned out that the ground state pairing gap of the Hubbard model obtained from this fRG + MFT approach agrees well with results from the much more involved flow equations with coupled interaction channels. The method was then extended to parameter regions where pairing coexists with N\'eel \cite{Wang2014} or spiral \cite{Yamase2016} antiferromagnetism. Coexistence of antiferromagnetism and $d$-wave pairing was also found in quantum cluster calculations at stronger interactions. \cite{Lichtenstein2000,Aichhorn2006, Capone2006,Kancharla2008,Zheng2016}

So far, fRG calculations of order parameters for the repulsive two-dimensional Hubbard model have relied on a static approximation, that is, the frequency dependences of the effective interaction and gap functions have been neglected. \cite{footnote_attrHM}
However, dynamical fRG flows in the symmetric regime (before reaching the critical scale $\Lambda_c$) have indicated a rather strong impact of the frequency dependence already at moderate interaction strengths. \cite{Husemann2012,Vilardi2017} The effective two-particle interactions generally develop a strong dependence on all three Matsubara frequencies, and the critical scale is enhanced compared to the static approximation. More importantly, it turned out that the frequency dependence leads to an expansion of the parameter regime where antiferromagnetism is the first instability (at the critical scale $\Lambda_c$), with rather weak pairing interactions at that scale. This cast some doubt on the robust pairing tendencies obtained in the static fRG.

In the present work we address this issue by using a dynamical extension of the fRG + MFT method with full frequency dependence to compute magnetic and pairing gap functions for the repulsive two-dimensional Hubbard model with a moderate interaction strength. While magnetism is indeed the leading instability in a broad doping range, we find that robust pairing with a sizable pairing gap emerges in coexistence with antiferromagnetism at energy scales below $\Lambda_c$ around optimal doping. The size of the gap is not reduced compared to the one obtained in a static fRG + MFT approximation. We also compute the superfluid phase stiffness which enables us to estimate the Kosterlitz-Thouless transition temperature for superconductivity as a function of doping.

Our paper is structured as follows. In Sec.~II we define the model and we present the fRG flow equations. Results for the gap functions as a function of doping and frequency, for the phase stiffness, and the Kosterlitz-Thouless temperature are presented and discussed in Sec.~III. A summary and conclusion in Sec.~IV closes the article.

%
\section{Formalism}
%
%
\subsection{Model}
The Hubbard model \cite{Montorsi1992} describes spin-$\frac{1}{2}$ lattice fermions with quantum mechanical hopping amplitudes and a local interaction. Its Hamiltonian has the form
\begin{equation}
 \mathcal{H} = \sum_{i,j,\sigma} t_{ij} \, 
 c^{\dagger}_{i,\sigma} c^{\phantom\dagger}_{j,\sigma} +
 U \sum_i n_{i,\up}n_{i,\down} ,
\end{equation}
where $c^{\dagger}_{i,\sigma}$ ($c^{\phantom\dagger}_{i,\sigma}$) creates (annihilates) a fermion on site $i$ with spin orientation $\sigma$ ($\up$ or $\down$). We consider the two-dimensional case on a square lattice with a repulsive interaction $U>0$ at finite temperatures $T$. We restrict the hopping matrix $t_{ij}$ to nearest and next-to-nearest neighbors, with amplitudes $-t$ and $-t'$, respectively. Fourier transforming $t_{ij}$ yields the dispersion relation
\begin{equation}
 \eps_{\mathbf{k}} =
 -2t \left( \cos{k_x} + \cos{k_y} \right) -4 t' \cos{k_x} \cos{k_y} .
\end{equation}
%
%
\subsection{Functional renormalization group}
We compute the gap functions for magnetism and superconductivity from a truncated fRG flow. The flow is defined by a successive integration of the fermionic fields, which is implemented via a flowing cutoff applied to the bare propagator. \cite{Metzner2012} Here we choose a smooth frequency cutoff of the form \cite{Husemann2009}
\begin{equation}
 G_0^\Lambda(\bs{k},\nu) =
 \frac{\nu^2}{\nu^2 + \Lambda^2} \, G_0(\bs{k},\nu),
\end{equation}
where $G_0(\bs{k},\nu) = \left[i\nu - (\eps_\bs{k} - \mu) \right]^{-1}$ is the bare fermion propagator as a function of the crystal momentum $\bs{k}$ and the fermionic Matsubara frequency $\nu$ (odd integer multiples of $\pi T$).
The flow parameter $\Lambda$ is reduced continuously from infinity to zero.

At low temperatures the flow needs to be divided in two qualitatively distinct regimes. For $\Lambda > \Lambda_c$ all symmetries of the bare Hamiltonian are conserved, while for $\Lambda < \Lambda_c$ the SU(2) spin symmetry, the U(1) charge symmetry, or both are spontaneously broken. The instabilities are signalled by divergencies of effective interactions at the critical scale $\Lambda_c$. The latter is non-zero for temperatures below a pseudocritical temperature $T^*$, which depends on the hopping amplitudes, the interaction strength, and the band filling.

In the symmetric regime we approximate the flow by a second order (one-loop) truncation for the effective two-particle interaction, discarding self-energy feedback and contributions from the three-particle interaction. All fluctuation channels (charge, magnetic, pairing) and the coupling between these channels are taken into account on equal footing. Fluctuation driven instabilities such as $d$-wave pairing from magnetic fluctuations are captured by this weak-coupling truncation.
In the symmetry-broken regime ($\Lambda < \Lambda_c$) we further simplify the flow by keeping only those interactions which generate the order parameters. The flow of these reduced interactions is decoupled, that is, each of them is determined by a single channel only. The flowing reduced interactions determine the flow of the gap functions, which appear as anomalous self-energy contributions. Their feedback on the flow of the interactions is crucial in the symmetry-broken regime \cite{Salmhofer2004} and is therefore taken into account.
The fusion of a complete one-loop flow for $\Lambda > \Lambda_c$ with a single-channel truncation for $\Lambda < \Lambda_c$ corresponds to a mean-field approximation with effective interactions extracted from the flow at the critical scale $\Lambda_c$.\cite{Wang2014}

%
\subsection{Symmetric regime}
The two-particle vertex
$V^\Lambda_{\sigma_1\sigma_2\sigma_3\sigma_4}(k_1,k_2,k_3,k_4)$
with $k_i = (\bs{k}_i,\nu_i)$ is generally a function of four momentum, frequency, and spin variables, where the labels 1 and 2 correspond to ingoing, the labels 3 and 4 to outgoing particles. Translation invariance in space and time implies momentum and frequency conservation, $k_1 + k_2 = k_3 + k_4$. We can thus drop the redundant variable $k_4$ in $V^\Lambda$. In case of SU(2) spin rotation invariance, all spin components of the vertex can be expressed by a single momentum and frequency dependent function $V^\Lambda(k_1,k_2,k_3)$ as \cite{Salmhofer2001}
\begin{eqnarray}
 V^\Lambda_{\sigma_1\sigma_2\sigma_3\sigma_4}(k_1,k_2,k_3,k_4) &=&
 V^\Lambda(k_1,k_2,k_3)
 \delta_{\sigma_1\sigma_3} \delta_{\sigma_2\sigma_4} \nonumber \\
 &-& V^\Lambda(k_2,k_1,k_3)
 \delta_{\sigma_1\sigma_4} \delta_{\sigma_2\sigma_3} . \hskip 5mm
\end{eqnarray}

To parametrize the momentum dependence of the vertex, we use the channel decomposition introduced by Husemann and Salmhofer, \cite{Husemann2009} where the vertex is expressed as a sum of the bare interaction and fluctuation induced effective interactions in the pairing, magnetic, and charge channels,
\begin{eqnarray} \label{eq:channeldec}
 V^\Lambda(k_1,k_2,k_3) &=& U - \phi^\Lambda_p(k_1,k_3;k_1+k_2) \nonumber \\
 &+& \phi^\Lambda_m(k_1,k_2;k_2-k_3) \nonumber \\
 &+& {\textstyle \frac{1}{2}} \phi^\Lambda_m(k_1,k_2;k_3-k_1) \nonumber \\
 &-& {\textstyle \frac{1}{2}} \phi_c^\Lambda(k_1,k_2;k_3-k_1) .
\end{eqnarray}
The dependence on the two fermionic momenta in each channel is more regular than the dependence on the bosonic linear combination in the last argument. We will neglect it completely in the magnetic and charge channel, where only relatively weak momentum dependences are expected, and approximate it by a constant and a $d$-wave term in the pairing channel, that is, 
\begin{align} 
 \label{eq:phi_p}
 \phi_p^\Lambda(k_1,k_3;q) &= \mathcal{S}^\Lambda_{\bs{q},\omega}(\nu_1,\nu_3)
 \nonumber \\
 &+ d_{\bs{k}_1-\frac{\bs{q}}{2}}d_{\bs{k}_3-\frac{\bs{q}}{2}}
 \mathcal{D}^\Lambda_{\bs{q},\omega}(\nu_1,\nu_3), \\
 \label{eq:phi_m}
 \phi_m^\Lambda(k_1,k_2;q) &= \mathcal{M}^\Lambda_{\bs{q},\omega}(\nu_1,\nu_2), \\
 \label{eq:phi_c}
 \phi_c^\Lambda(k_1,k_2;q) &= \mathcal{C}^\Lambda_{\bs{q},\omega}(\nu_1,\nu_2) ,
\end{align}
with $q = (\bs{q},\omega)$ and $d_{\bs{k}} = \cos k_x-\cos k_y$. The dependence on the remaining bosonic momentum $\bs{q}$ is kept, and the dependence on all three Matsubara frequencies $\nu_1$, $\nu_2$, and $\omega$ is fully taken into account.
The parametrization of $V^\Lambda(k_1,k_2,k_3)$ via Eqs.~(\ref{eq:channeldec}) - (\ref{eq:phi_c}) has already been used in Ref.~\onlinecite{Vilardi2017}. The flow equations for the four functions $\mathcal{S}^\Lambda$, $\mathcal{D}^\Lambda$, $\mathcal{M}^\Lambda$, and $\mathcal{C}^\Lambda$ can be found there.
%
%
\subsection{Symmetry-broken regime}
In the regime $\Lambda < \Lambda_c$ at least one symmetry of the Hubbard Hamiltonian is spontaneously broken. The flow of the effective interactions \cite{Metzner2012} and most studies by other methods \cite{Scalapino2012} indicate antiferromagnetism (N\'eel, stripes, spiral etc.) and $d$-wave pairing as the key instabilities. We restrict the zoo of possible magnetic order patterns to spiral order with a single wave vector $\bQ$, which includes N\'eel order as the special case where $\bQ = (\pi,\pi)$. Spiral order is planar and can be oriented in any plane. The most convenient choice is a plane perpendicular to the spin-quantization axis, that is, the $xy$-plane for the standard spin basis.

Spiral magnetic order in the $xy$-plane is associated with anomalous expection values
$\langle \psi_\up^{\phantom *}(k) \psi_\down^*(k+Q) \rangle$, where $Q = (\bQ,0)$,
and singlet pairing with anomalous expectation values of the form
$\langle \psi_\up(k) \psi_\down(-k) \rangle$.
Here and in the following $\psi_\sigma^*(k)$ and $\psi_\sigma(k)$ are Grassmann fields corresponding to fermion creation and annihilation operators in momentum representation, respectively.
Symmetry breaking leads to anomalous terms in the effective action $\Gamma^\Lambda[\psi,\psi^*]$. Spiral order leads to spin-flips combined with a momentum shift $\bQ$, and pairing leads to pair creation and annihilation terms. The quadratic part of the effective action thus has the general form
\begin{eqnarray} \label{eq:Gamma_2}
 \Gamma_2^\Lambda[\psi,\psi^*] &=&
 \int_k \sum_\sigma \left[ - (G_0^\Lambda(k))^{-1} + \Sigma^\Lambda(k) \right] \, 
 \psi_\sigma^*(k) \psi_\sigma(k) \nonumber \\
 &+& \int_k \left[
 \Delta_m^\Lambda(k) \, m^*(k) + \Delta_m^{\Lambda *}(k^*) \, m(k) \right] \nonumber \\
 &+& \int_k \left[
 \Delta_p^\Lambda(k) \, p^*(k) + \Delta_p^{\Lambda *}(k^*) \, p(k) \right] ,
\end{eqnarray}
where $k^* = (\bk,-\nu)$, 
\begin{align}
 m(k) &= \psi_\up^{\phantom *}(k) \, \psi_\down^*(k+Q) , \quad
 m^*(k) = \psi_\down^{\phantom *}(k+Q) \, \psi_\up^*(k) , \nonumber \\
 p(k) &= \psi_\up(k) \, \psi_\down(-k) , \hskip 9.5mm
 p^*(k) = \psi_\down^*(-k) \, \psi_\up^*(k) , \nonumber
\end{align}
and $\int_k = T \sum_\nu \int_\bk$ with $\int_\bk = \int \frac{d^2\bk}{(2\pi)^2}$ is a shorthand notation for momentum integrals and Matsubara frequency sums.
$\Sigma^\Lambda(k)$ is the normal self-energy, which we neglect in this work.
$\Delta_m^\Lambda(k)$ and $\Delta_p^\Lambda(k)$ are (generally) complex functions, which we refer to as magnetic gap function and pairing gap function, respectively.
The frequency dependence of the (spin-singlet) pairing gap is symmetric, that is, $\Delta_p^\Lambda(k) = \Delta_p^\Lambda(k^*)$.

It is convenient to express $\Gamma_2^\Lambda[\psi,\psi^*]$ in a Nambu representation as
\begin{equation}
 \Gamma_2^\Lambda[\Psi,\Psi^*] =
 - \int'_k \Psi^*(k) \left[ \mathcal{G}^\Lambda(k) \right]^{-1} \Psi(k) \, ,
\end{equation}
with a 4-component Nambu spinor
\begin{equation}
 \Psi(k) =
 \left[ \psi_\up(k),\psi_\down^*(-k),\psi_\down(k+Q),\psi_\up^*(-k-Q)
 \right] \, ,
\end{equation}
and a $4\times4$ Nambu propagator
$\mathcal{G}^\Lambda(k) = - \langle \Psi(k) \Psi^*(k) \rangle$.
The prime at the integral indicates that the momentum integration is restricted to a reduced magnetic Brillouin zone. Its shape depends on $\bQ$.
For example, for $\bQ = (\pi - 2\pi\eta,\pi)$, a suitable reduced Brillouin zone is given by $\left\{ \bs{k}; |k_x| \leq \pi, |k_y| \leq \pi/2 \right\}$.
The matrix elements of the inverse Nambu propagator are determined by Eq.~(\ref{eq:Gamma_2}) as
\begin{widetext}
\begin{equation} \label{eq:Nambuprop}
 \left[ \mathcal{G}^\Lambda(k) \right]^{-1} = 
 \left( 
 \begin{array}{cccc}
 [G_0^\Lambda(k)]^{-1} & \Delta_p^\Lambda(k) & \Delta_m^{\Lambda}(k) & 0
 \\[2mm]
 \Delta_p^{\Lambda *}(k^*) & - [G_0^\Lambda(-k)]^{-1} & 0 & - \Delta_m^{\Lambda}(-k-Q)
 \\[2mm]
 \Delta_m^{\Lambda *}(k^*) & 0 & [G_0^\Lambda(k+Q)]^{-1} & - \Delta_p^\Lambda(-k-Q)
 \\[2mm]
 0 & - \Delta_m^{\Lambda *}(-k^*-Q^*) & - \Delta_p^{\Lambda *}(-k^*-Q^*) &
  - [G_0^\Lambda(-k-Q)]^{-1}
 \end{array} \right) \, .
\end{equation}
\end{widetext}

Our central approximation in the regime $\Lambda < \Lambda_c$ is that we keep only those effective interactions which contribute {\em directly}\/ to the flow of the gap functions.
The corresponding reduced quartic part of the effective action has the form
\begin{eqnarray} \label{eq:Gamma_4}
 \Gamma_4^\Lambda[\psi,\psi^*] \! &=& \!\!
 \int_{k,k'} \!\!\! \frac{V_m^\Lambda(k,k')}{2} \,
 [m^*(k) m(k') + m(k) m^*(k')] \hskip 3mm \nonumber \\ \! &+& \!\!
 \int_{k,k'} \!\!\! \frac{W_m^\Lambda(k,k')}{2} \,
 [m^*(k) m^*(k') + m(k) m(k')] \nonumber \\ \! &+& \!\!
 \int_{k,k'} \!\!\! \frac{V_p^\Lambda(k,k')}{2} \,
 [p^*(k) p(k') + p(k) p^*(k')] \nonumber \\ \! &+& \!\!
 \int_{k,k'} \!\!\! \frac{W_p^\Lambda(k,k')}{2} \,
 [p^*(k) p^*(k') + p(k) p(k')] ,
\end{eqnarray}
The coupling functions $V_m^\Lambda(k,k')$ and $V_p^\Lambda(k,k')$ para\-metrize {\em reduced} normal interactions. For example, $V_p^\Lambda(k,k')$ corresponds to the coupling function of the reduced BCS model, which is restricted to the Cooper channel (with vanishing total momentum of ingoing and outgoing particles).
$W_m^\Lambda(k,k')$ and $W_p^\Lambda(k,k')$ para\-metrize anomalous interaction terms which are generated in the symmetry broken regime. \cite{footnote_W}

At the critical scale $\Lambda_c$ (and above it), the anomalous terms vanish, while the normal reduced coupling functions can be obtained from the full two-particle vertex
$V_{\sigma_1\sigma_2\sigma_3\sigma_4}^{\Lambda}(k_1,k_2,k_3,k_4)$ as \cite{footnote_Yamase}
\begin{eqnarray}
 V_m^{\Lambda}(k,k') &=&
 V_{\sigma,-\sigma,-\sigma,\sigma}^{\Lambda}(k+Q,k',k,k'+Q) , \\
 V_p^{\Lambda}(k,k') &=&
 \frac{1}{2} V_s^{\Lambda}(k,-k,k',-k') ,
\end{eqnarray}
where $V_s^{\Lambda} =
V_{\sigma,-\sigma,\sigma,-\sigma}^\Lambda - V_{\sigma,-\sigma,-\sigma,\sigma}^\Lambda$
is the spin-singlet component of the two-particle vertex.
Note that $V_m^{\Lambda}(k,k')$ is generally complex, while $V_p^{\Lambda}(k,k')$ is real, since the imaginary parts cancel in the spin-singlet component.

Due to the restrictions of momenta in the reduced interactions, the flows of the magnetic and pairing coupling functions are decoupled from each other, and each of them is governed by one channel only. For example, the flow of the pairing coupling is determined by the particle-particle channel.
The flow of the gap functions is entirely determined by the {\em amplitude}\/ coupling \cite{Salmhofer2004,Gersch2008,Eberlein2010}
\begin{equation}
 \mathcal{A}_X^\Lambda(k,k') = V_X^\Lambda(k,k') + W_X^\Lambda(k,k') ,
\end{equation}
with $X = m,p$, so that we consider only this linear combination in the following.
The {\em transverse} coupling $V_X^\Lambda(k,k') - W_X^\Lambda(k,k')$ is related to a Ward identity and the Goldstone theorem, which are fulfilled in the fRG + MFT approach. \cite{Bonetti2020}

In line with our parametrization in the symmetric regime, we simplify the momentum dependencies of the gap functions and the coupling functions by using a small set of form factors. In the magnetic channel, the dependences on $\bk$ and $\bk'$ are weak and will be neglected. In the pairing channel we keep only the $d$-wave components, since there is no pairing instability with any other symmetry.
Hence, we approximate the momentum dependences of the gap and coupling functions by a simple ansatz, namely
\begin{eqnarray}
    \Delta_m^\Lambda(k) &=& \Delta_m^\Lambda(\nu) , \\
    \Delta_p^\Lambda(k) &=& \Delta_p^\Lambda(\nu) \, d_\bs{k} ,
\end{eqnarray}
for the gap functions, and 
\begin{eqnarray}   
    \mathcal{A}_m^\Lambda(k,k') &=& \mathcal{A}_m^\Lambda(\nu,\nu') , \\
    \mathcal{A}_p^\Lambda(k,k') &=& \mathcal{A}_p^\Lambda(\nu,\nu') \,
    d_{\bs{k}} d_{\bs{k}'}.
\end{eqnarray}
for the coupling functions. The dependences on the Matsubara frequencies $\nu$ and $\nu'$ are fully taken into account.

At the critical scale $\lamc$, we have
$\Delta_m^{\lamc}(\nu) = \Delta_p^{\lamc}(\nu) = 0$, and 
\begin{eqnarray}
    \mathcal{A}_m^{\lamc}(\nu,\nu') &=& V_m^{\lamc}(\nu,\nu') =
    \int_{\bk,\bk'} V_m^{\lamc}(k,k') , \\
    \mathcal{A}_p^{\lamc}(\nu,\nu') &=& V_p^{\lamc}(\nu,\nu') =
    \int_{\bk,\bk'} d_{\bk} d_{\bk'} V_p^{\lamc}(k,k') . \hskip 5mm
\end{eqnarray}
These are the initial conditions for the flow in the symmetry-broken regime $\Lambda < \lamc$. In our numerical solution we will switch from the full one-loop flow in the symmetric regime to the reduced single-channel flow slightly above the critical scale $\lamc$, and we insert tiny gap values as initial condition for the gap functions to get the symmetry breaking started.

Since only a single channel contributes to the flow of each coupling function $\mathcal{A}_X^\Lambda$, the right hand side of the flow equation is a quadratic form in $\mathcal{A}_X^\Lambda$.
It is convenient to view $\mathcal{A}_X^\Lambda$ as a matrix with matrix elements $\mathcal{A}_X^\Lambda(\nu,\nu')$. The flow equation for $\mathcal{A}_X^\Lambda$ can then be written in matrix form,
\begin{equation} \label{eq:flowA_X}
 \partial_\Lambda \mathcal{A}_X^\Lambda = \mathcal{A}_X^\Lambda
 \left[\partial_\Lambda\Pi_X^\Lambda\right] \mathcal{A}_X^\Lambda ,
\end{equation}
for $X = m,p$, where $\Pi_X^\Lambda(\nu,\nu') = \delta_{\nu\nu'} \Pi_X^\Lambda(\nu)$ is diagonal in frequency with diagonal elements
\begin{eqnarray}
 \label{eq:Pi_m}
 \Pi_m^\Lambda(\nu) &=& T \int_\bs{k}
 \left[ G^\Lambda(k) G^\Lambda(k+Q) + (F_m^\Lambda(k))^2 \right] , \\
 \label{eq:Pi_p}
 \Pi_p^\Lambda(\nu) &=& T \int_\bs{k} d_\bs{k}^2
 \left[-G^\Lambda(k) G^\Lambda(-k) + (F_p^\Lambda(k))^2 \right] . \hskip 5mm
\end{eqnarray}
The propagators in Eqs.~(\ref{eq:Pi_m}) and (\ref{eq:Pi_p}) are defined by the expectation values
\begin{eqnarray}
 G^\Lambda(k) &=& - \langle \psi_\sigma(k) \psi_\sigma^*(k) \rangle =
 \mathcal{G}_{11}^\Lambda(k) , \\[1mm]
 F_m^\Lambda(k) &=& - \langle \psi_\up(k) \psi_\down^*(k+Q) \rangle =
 \mathcal{G}_{13}^\Lambda(k) , \\[1mm]
 F_p^\Lambda(k) &=& - \langle \psi_\up(k) \psi_\down(-k) \rangle =
 \mathcal{G}_{12}^\Lambda(k) ,
\end{eqnarray}
where $\mathcal{G}_{\alpha\alpha'}^\Lambda(k)$ are matrix elements of the Nambu propagator defined in Eq.~(\ref{eq:Nambuprop}).

The flow equation (\ref{eq:flowA_X}) can be formally integrated. With the initial condition $\mathcal{A}_X^\Lambda = \mathcal{A}_X^\lamc = V_X^\lamc$ for $\Lambda = \lamc$, one obtains the solution
\begin{equation} \label{eq:A_X}
 \mathcal{A}_X^\Lambda =
 \left[ 1 - \tilde V_X^{\lamc} \Pi_X^\Lambda \right]^{-1}
 \tilde V_X^{\lamc} ,
\end{equation}
where $[\dots]^{-1}$ denotes a matrix inversion, and
\begin{equation} \label{eq:Irr}
 \tilde V_X^{\lamc} = \left[ 1 +
 V_X^{\lamc} \Pi_X^\lamc \right]^{-1} V_X^{\lamc} .
\end{equation}
$\tilde V_X^{\Lambda_c}$ and $V_X^{\Lambda_c}$ are related by the Bethe-Salpeter equation
$V_X^{\Lambda_c} = \tilde V_X^{\Lambda_c} + 
 \tilde V_X^{\Lambda_c} \Pi_X^\lamc V_X^{\Lambda_c}$.
Hence, $\tilde V_X^{\Lambda_c}$ is the two-particle {\em irreducible}\/ part of $V_X^{\Lambda_c}$.

The flow equations for the gap functions are obtained from the first equation in the flow equation hierarchy, which relates the flow of the self-energy to the two-particle vertex. \cite{Metzner2012} Inserting the two-particle vertex as described above, one finds
\begin{eqnarray}
 \label{eq:flowDelta_m}
 \partial_{\Lambda} \Delta_m^\Lambda(\nu) &=& 
 - T \sum_{\nu'} \int_\bs{k} \mathcal{A}_m^\Lambda(\nu,\nu') \,
 \widetilde{\partial}_\Lambda F_m^\Lambda(\bs{k},\nu') , \\
 \label{eq:flowDelta_p}
 \partial_{\Lambda} \Delta_p^\Lambda(\nu) &=& 
 - T \sum_{\nu'} \int_\bs{k} \mathcal{A}_p^\Lambda(\nu,\nu') \,
 \widetilde{\partial}_\Lambda F_p^\Lambda(\bs{k},\nu') \, d_\bk , \hskip 5mm
\end{eqnarray}
where $F^\Lambda_m(k)$ and $F^\Lambda_p(k)$ are the anomalous propagators corresponding to magnetic order and pairing, respectively.
The derivative denoted by $\widetilde{\partial}_\Lambda$ acts only on the scale dependence directly introduced by the cutoff function, not on the scale dependence from the flowing gap functions appearing in the expressions for $F^\Lambda_m(k)$ and $F^\Lambda_p(k)$.
By contrast, the scale derivative $\partial_\Lambda$ in Eq.~(\ref{eq:flowA_X}) is a total derivative, which includes self-energy feedback terms generated from tadpole contractions of three-particle vertices. \cite{Katanin2004,Salmhofer2004}

The flow equations (\ref{eq:flowDelta_m}) and (\ref{eq:flowDelta_p}) can be formally integrated to \cite{Wang2014}
\begin{eqnarray}
 \label{eq:Delta_m}
 \Delta_m^\Lambda(\nu) &=& 
 - T \sum_{\nu'} \int_\bs{k} \tilde V_m^\lamc(\nu,\nu') \,
 F_m^\Lambda(\bs{k},\nu') , \\
 \label{eq:Delta_p}
 \Delta_p^\Lambda(\nu) &=& 
 - T \sum_{\nu'} \int_\bs{k} \tilde V_p^\lamc(\nu,\nu') \,
 F_p^\Lambda(\bs{k},\nu') \, d_\bk , \hskip 5mm
\end{eqnarray}
where $\tilde V_X^\lamc$ is the irreducible part of $V_X^\lamc$.
These are non-linear integral equations for the gap functions $\Delta_X^\Lambda(\nu)$. Finding a self-consistent solution by iteration is difficult. We found that it is much easier to compute the gap functions from a numerical integration of the flow equations (\ref{eq:flowDelta_m}) and (\ref{eq:flowDelta_p}).

The pairing gap function is symmetric in frequency, that is, $\Delta_p(\nu) = \Delta_p(-\nu)$, and can be chosen real for all frequencies, since $\tilde V_p^\lamc(\nu,\nu')$ is real for all frequencies. The magnetic gap function can be chosen such that $\Delta_m(-\nu) = \Delta_m^*(\nu)$. Its imaginary part cannot be removed by choosing a suitable global phase, since the effective interaction $\tilde V_m^\lamc(\nu,\nu')$ is complex.


\section{Results}

Before presenting our results, we mention here a few technical details regarding the numerical solution of the equations.
In the symmetric regime, we solve the flow equation of the competing channels by taking into account about 90 Matsubara frequencies for each frequency argument and about 320 Brillouin zone patches for the transfer momentum. When required, we extend the frequency range with a numerical projection. \cite{Wentzell2016} We have checked that the frequency boxes are large enough to obtain converged results.
The symmetric one-loop flow is stopped when the effective interaction in one of the channels reaches the value $400t$, that is, at a scale $\Lambda$ very close to the critical scale $\Lambda_c$.
In the symmetry-broken regime, the two order parameters are initialized with a small value of the order $10^{-3}t$.
The irreducible vertices, Eq.~(\ref{eq:Irr}), and the flowing vertices, Eq.~(\ref{eq:A_X}), in the symmetry-broken regime are calculated with a matrix inversion in Matsubara frequency space. 
The momentum $\bs{Q}$ characterizing the spiral order is determined from the maximum of the magnetic interaction upon approaching the critical scale $\Lambda_c$ in the symmetric regime. It has the general form $\bs{Q} = (\pi - 2\pi\eta,\pi)$, or symmetry-related. We have checked that the peak momentum of the magnetic interaction at $\Lambda_c$ is very close to the momentum $\bQ$ which minimizes the free energy at the end of the flow.

We choose a fixed hopping amplitude ratio $t'/t = -0.16$ and a moderate interaction strength $U = 3t$ in all calculations. We use natural units such that $\hbar = k_B = 1$. The lattice constant is also set to one. Quantities with dimension energy (= temperature) are presented in units of $t$.


\subsection{Order parameters}

\begin{figure}[t!]
 \includegraphics[scale=0.52]{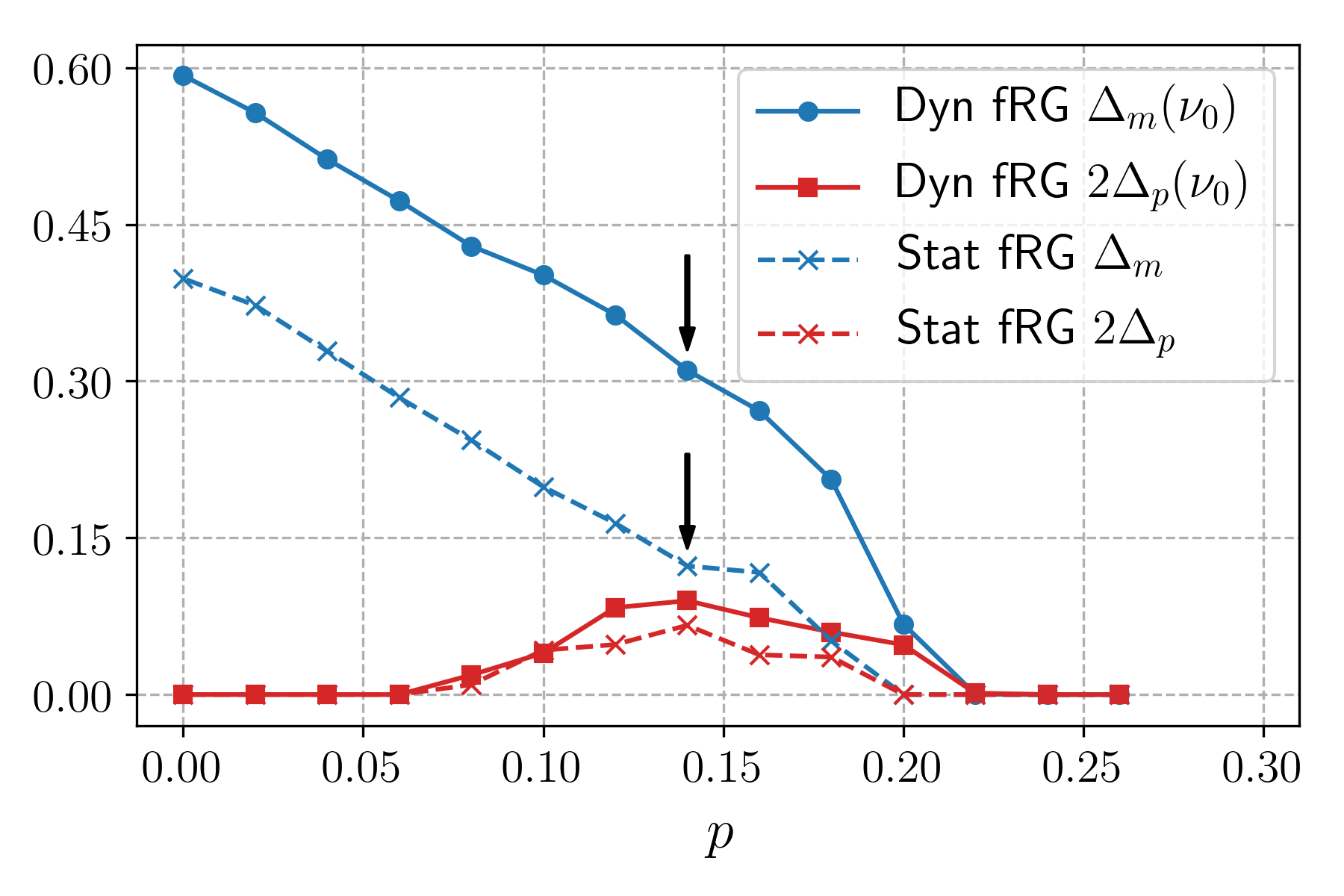}
 \caption{Amplitudes of the magnetic and pairing gaps as a function of doping
 at fixed temperature $T=0.027t$. The gaps from the dynamical fRG with full frequency
 dependence are shown for the lowest Matsubara frequency $\nu_0 = \pi T$, and
 compared to results from the static approximation. The vertical arrows indicate
 the transition from N\'eel to incommensurate spiral antiferromagnetism.}
 \label{fig:gaps}
\end{figure}
In Fig.~\ref{fig:gaps}, we show the amplitudes (maxima in momentum space) of the magnetic and the pairing gaps as a function of the hole doping $p=1-n$ at a fixed temperature $T=0.027t$. Results from the dynamical fRG with full frequency dependence as described above are compared to results from a static approximation as employed by Wang {\em et al.} \cite{Wang2014} and Yamase {\em et al.}, \cite{Yamase2016} where all frequency dependences are neglected.
The gap amplitudes obtained from the dynamical fRG are shown at the lowest positive Matsubara frequency, that is, $\nu_0 = \pi T$. The pairing gap amplitude is $2\Delta_p(\nu)$, since the maximum value of the $d$-wave form factor $d_\bk = \cos k_x - \cos k_y$ is two. Within our ansatz, the magnetic gap is momentum independent. Its ``amplitude'' is thus simply $\Delta_m(\nu)$. It is generally complex, but the imaginary part is small for $\nu = \nu_0 = \pi T$. The imaginary part of the analytic continuation of $\Delta_m(\nu)$ to the entire complex frequency plane vanishes for $\nu \to 0$. In the figure we only show the real part of $\Delta_m(\nu_0)$. In the static approximation the magnetic gap is frequency independent and can be chosen real.

Antiferromagnetic order extends from half-filling to a doping value of around $20\%$, with N\'eel order up to about $13\%$, and incommensurate spiral order beyond. The transition between N\'eel order and incommnsurate spiral order is discontinuous, with a pronounced jump of the incommensurability $\eta$.
In the range between $8\%$ and $20\%$ doping, a sizable pairing order parameter appears. Hence, our calculation confirms the coexistence of antiferromagnetism and $d$-wave pairing in the two-dimensional Hubbard model, as previously obtained from static fRG calculations. \cite{Reiss2007,Wang2014,Yamase2016}.
The doping dependence of the incommensurability $\eta$ also agrees well with previous static fRG calculations. \cite{Yamase2016}
The pairing mechanism is magnetic, that is, the attraction in the $d$-wave pairing channel is predominantly generated by antiferromagnetic fluctuations. The suppression of the pairing amplitude close to half-filling is caused by the magnetic gap leading to a truncation of the Fermi surface to small hole pockets. 

The gap amplitudes obtained from the static fRG are somewhat smaller than those obtained from the dynamical fRG, but they exhibit a similar qualitative doping dependence.
The magnetic order in the static approximation is weaker for two reasons, which have been revealed already previously. \cite{Vilardi2017} First, there is a minimum of the magnetic effective interaction $\mathcal{M}^\Lambda_{\bQ,\omega=0}(\nu_1,\nu_2)$ at the lowest fermionic Matsubara frequencies $|\nu_i| = \pi T$. In the static approximation this minimum value is practically extended to all frequencies $\nu_1$ and $\nu_2$. Second, in the static approximation the suppression of magnetic interactions from other fluctuation channels is overestimated. \cite{Vilardi2017}
The pairing gap is also reduced in the static approximation. Since the pairing mechanism is mostly magnetic, one reason for this reduction is certainly the weaker magnetic interaction. On the other hand, the decay of the effective pairing interaction at large frequencies is neglected in the static approximation, leading to an enhancement of pairing tendencies. \cite{Husemann2012,Vilardi2017} The net effect seems to be a moderate reduction of the pairing gap by the static approximation.
Overall, inspite of strong frequency dependences of the interaction vertex,  \cite{Husemann2012,Vilardi2017} the static approximation does not entail a major error in the size of the magnetic and pairing gaps.

The influence of the frequency dependence of the magnetic interaction vertex on pairing was previously analyzed by Kitatani et al. \cite{Kitatani2018} at an intermediate coupling strength ($U=6t$) within the dynamical vertex approximation. \cite{Toschi2007}
In agreement with our results, they found a minimum at low frequencies. They concluded that this minimum leads to a significant reduction of the energy scale for pairing. This is not in conflict with our results, because they compared their result to that from a simple random phase approximation for the magnetic interaction, which grossly overestimates its strength in the relevant frequency range.

\begin{figure}[t!]
 \includegraphics[scale=0.52]{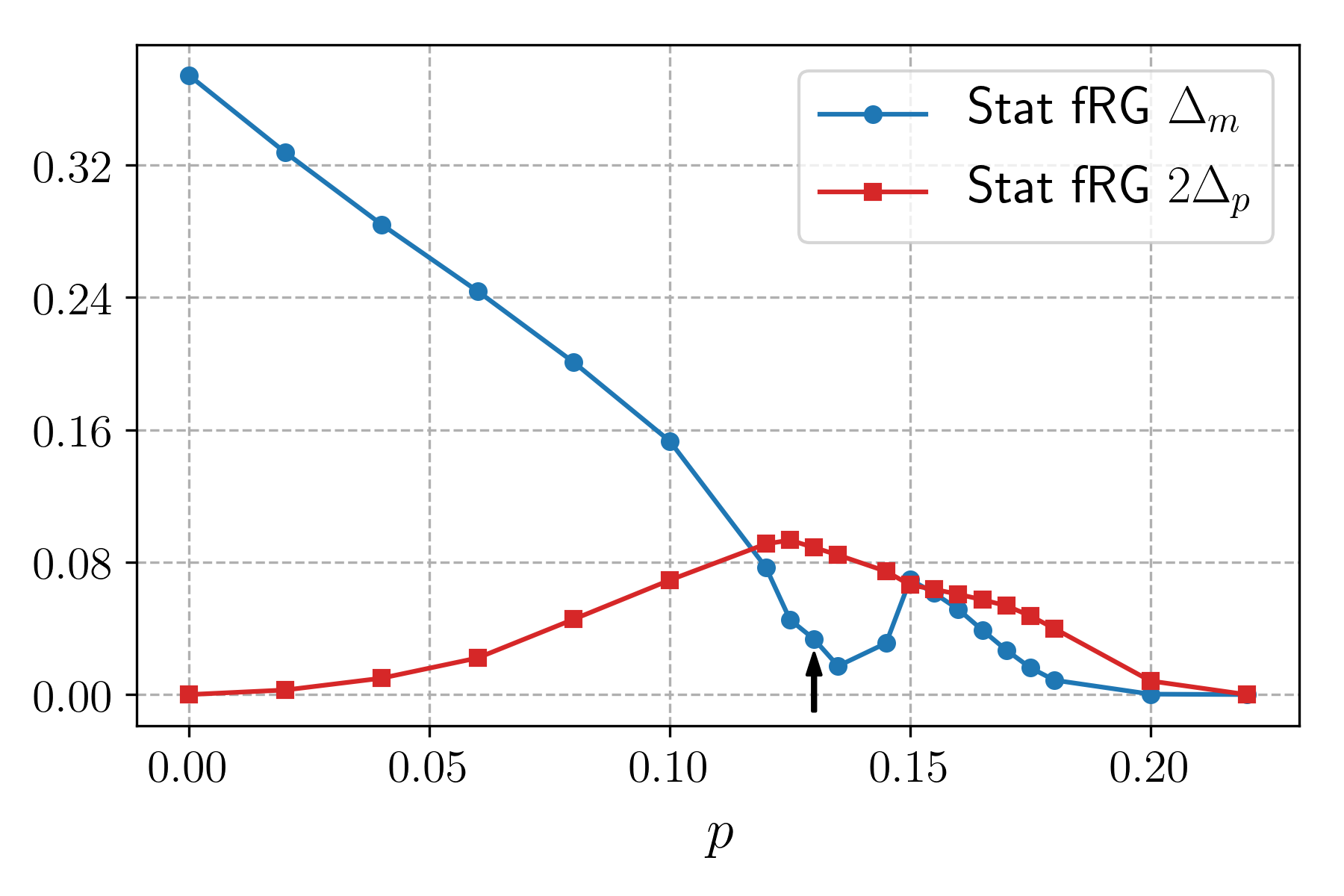}
 \caption{Amplitudes of the magnetic and pairing gap as a function of doping as obtained
 from the static fRG at the temperature $T=0.006t$. The vertical arrow indicates
 the transition from N\'eel to incommensurate spiral antiferromagnetism.}
 \label{fig:gaps_static}
\end{figure}
Yamase et al. \cite{Yamase2016} observed a pronounced dip of the magnetic gap at van Hove filling, where the pairing gap suppresses the magnetic order completely.
This feature is not visible in Fig.~\ref{fig:gaps}, neither in the static nor in the dynamical approximation. The discrepency is probably due to the finite temperature in our calculation. At present, we can access lower temperatures only by neglecting frequency dependences. In Fig.~\ref{fig:gaps_static} we show the doping dependence of the gap amplitudes at $T = 0.006t$ as obtained from the static fRG. Here, a dip of the magnetic gap at van Hove filling ($p_{\rm vH} \approx 14\%$ for our parameters) is clearly visible. We expect that the dip will become even more pronounced upon further lowering the temperature. 


\subsection{Flow and frequency dependence}

\begin{figure}[t!]
 \includegraphics[scale=0.55]{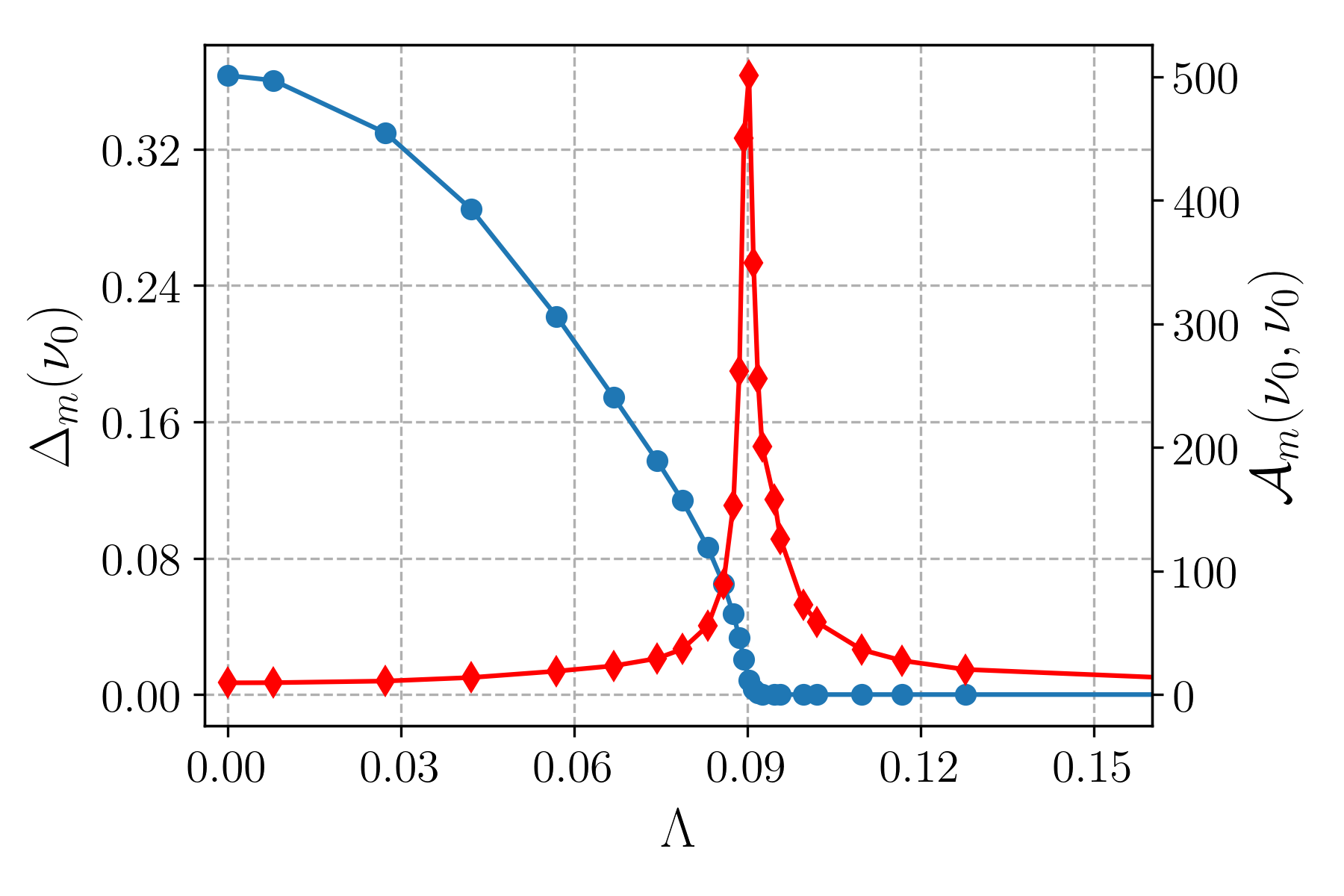}
 \caption{Flow of the magnetic gap $\Delta_m^\Lambda(\nu_0)$ and of the magnetic
 amplitude coupling $\mathcal{A}_m^\Lambda(\nu_0,\nu_0)$ with $\nu_0 = \pi T$
 at doping $p=0.12$ and temperature $T=0.027t$.}
 \label{fig:DeltaFlowMag}
\end{figure}
In this section we present some details on the flow and on the frequency dependences for a fixed doping $p = 0.12$ and a fixed temperature $T = 0.027t$.
In Fig.~\ref{fig:DeltaFlowMag} we show the flow of the magnetic gap $\Delta_m^\Lambda(\nu_0)$ at the lowest positive Matsubara frequency $\nu_0 = \pi T$ together with the flow of the magnetic amplitude coupling $\mathcal{A}_m^\Lambda(\nu_0,\nu_0)$. The flow looks qualitatively the same as for the BCS model. \cite{Salmhofer2004} The effective interaction increases rapidly upon approaching the critical scale $\Lambda_c$ from above and decreases again below it, approaching eventually a moderate finite value. The peak at $\Lambda_c$ is regularized by the small initial gap term inserted by hand. The gap increases monotonically from its tiny initial value at $\Lambda_c$ to a much larger value for $\Lambda \to 0$.
The flows of $\Delta_m^\Lambda(\nu)$ and $\mathcal{A}_m^\Lambda(\nu,\nu')$ have the same qualitative behavior for all choices of the Matsubara frequencies $\nu$ and $\nu'$.
The flow of the pairing gap and pairing coupling looks similar, too, but the onset of the gap and the peak in the coupling is situated at a lower scale. 

\begin{figure}[t!]
 \includegraphics[scale=0.55]{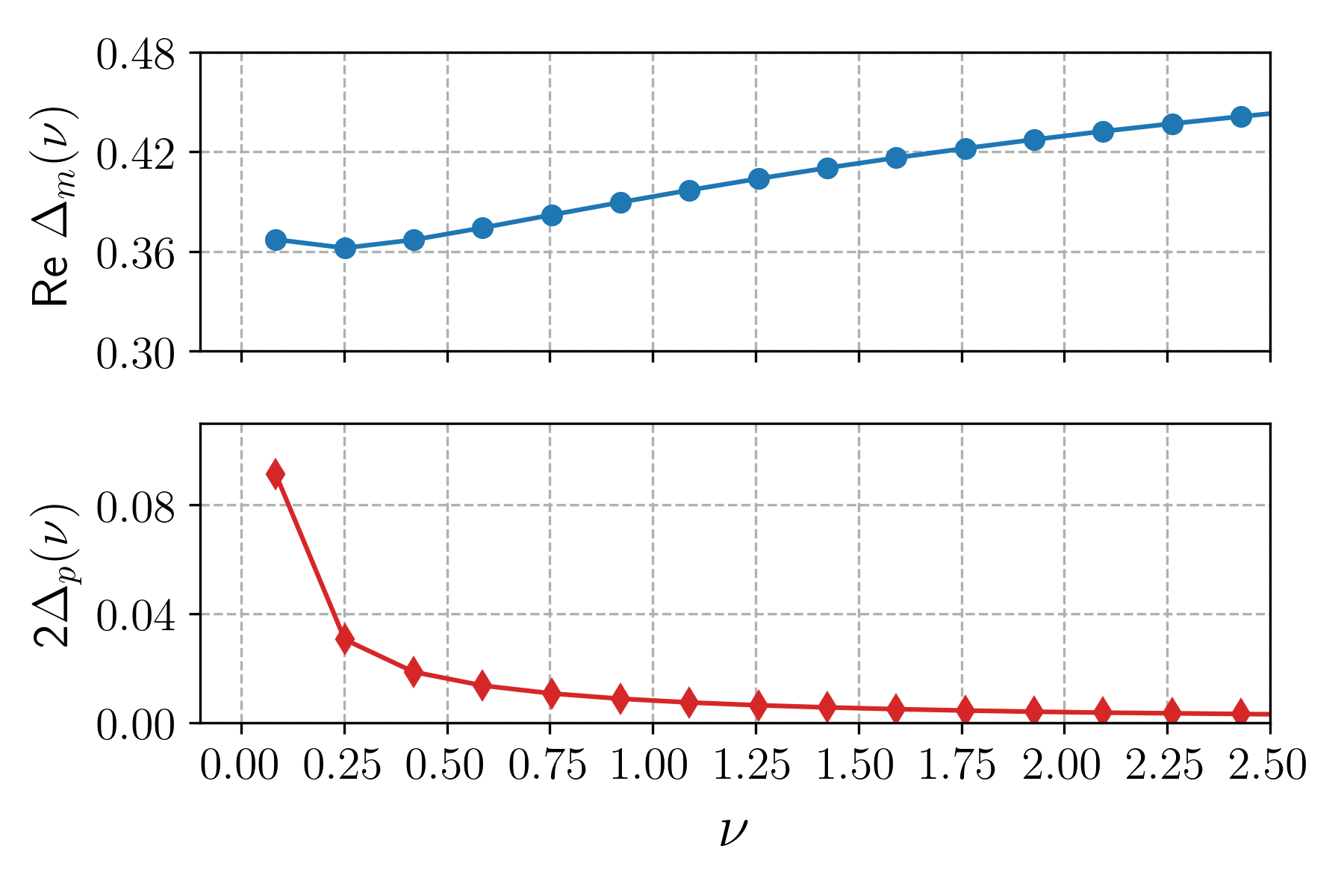}
 \caption{Real part of the magnetic gap amplitude and pairing gap amplitude
 as a function of frequency for $p=0.12$ and $T=0.027t$.}
 \label{fig:DeltaFreq}
\end{figure}
In Fig.~\ref{fig:DeltaFreq} we show the frequency dependence of the gap amplitudes at the end of the flow. The pairing gap can be chosen real for all frequencies, while the magnetic gap is necessarily complex. It is obvious that the frequency dependence of ${\rm Re} \Delta_m(\nu)$ is very weak, while $\Delta_p(\nu)$ decays rather quickly over the first few Matsubara frequencies. Hence, an accurate continuation of the pairing gap to zero frequency is difficult.

\begin{figure}[t!]
 \vskip 5mm
 \includegraphics[scale=0.45]{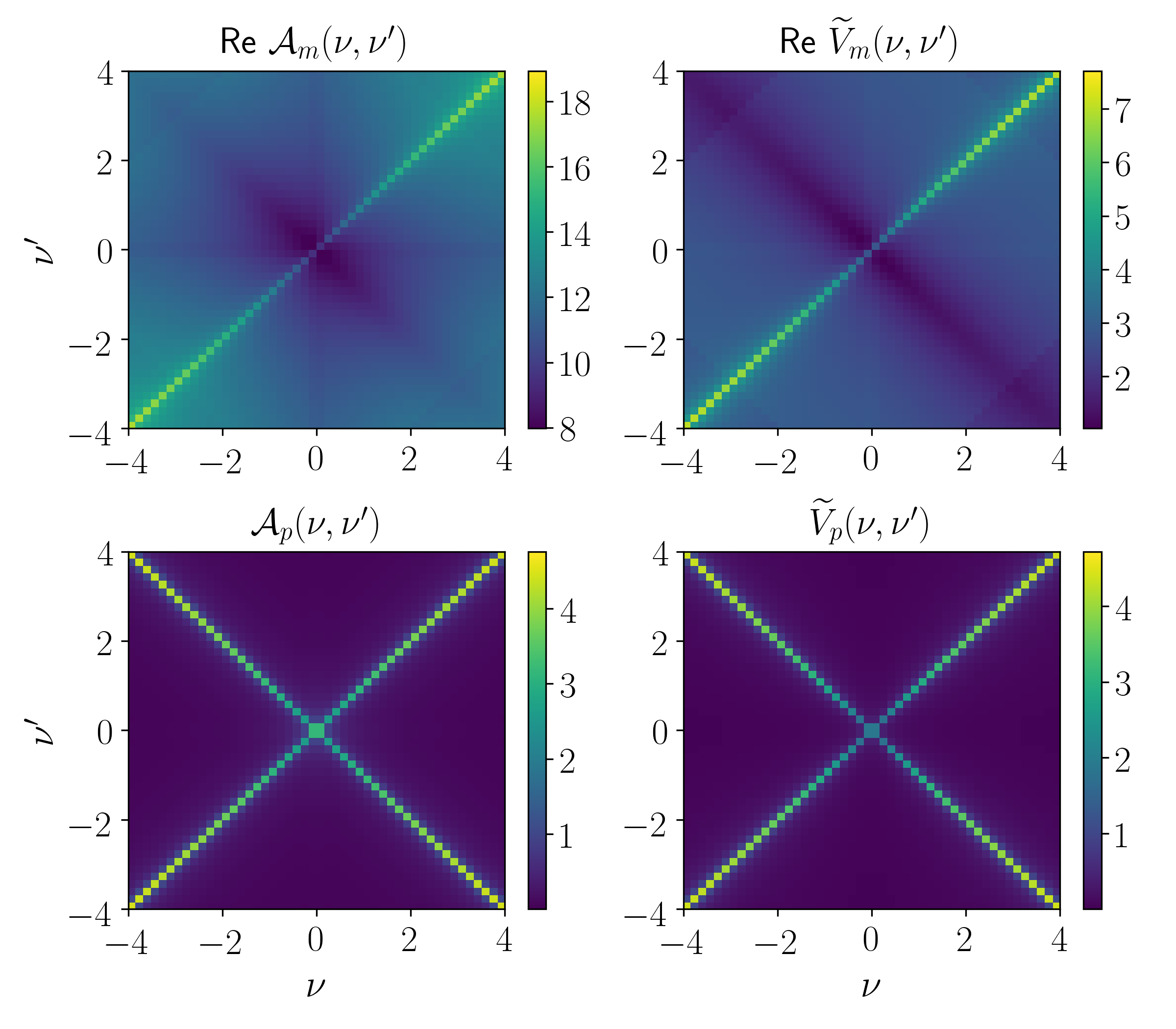}
 \caption{Frequency dependence of effective interactions in the magnetic (top) and pairing channels (bottom) at doping $p=0.12$ and temperature $T=0.027t$. The amplitude coulings $\mathcal{A}_X^\Lambda(\nu,\nu')$ at the end of the flow ($\Lambda \to 0$) are compared to the two-particle irreducible interaction parts $\tilde V_X^{\Lambda_c}(\nu,\nu')$ at the critical scale $\Lambda_c$. The effective interactions in the magnetic channel have a small imaginary part which is not shown.}
 \label{fig:VertexFreq}
\end{figure}
The frequency dependence of effective magnetic and pairing interactions is shown in Fig.~\ref{fig:VertexFreq}. There is a pronounced structure on the diagonal $\nu' = \nu$ in the magnetic channel, and for $\nu' = \pm \nu$ in the pairing channel. Note that these features are not present in the coupling functions $\mathcal{M}_{\bq,\omega}^\Lambda(\nu,\nu')$ and $\mathcal{D}_{\bq,\omega}^\Lambda(\nu,\nu')$, respectively, but are rather due to peaks of the other coupling functions as a function of the bosonic frequency $\omega$ at $\omega = 0$.
For the magnetic interaction, a qualitatively similar frequency structure is also obtained from a dynamical mean-field approximation, which neglects all correlations except the local ones. \cite{Rohringer2012}
The frequency dependence is obviously generated in the symmetric regime, so that it is fully developed already at the critical scale $\Lambda_c$. 
The amplitude coulings $\mathcal{A}_X^\Lambda(\nu,\nu')$ at the end of the flow ($\Lambda \to 0$) exhibit the same frequency dependence as the two-particle irreducible interaction parts $\tilde V_X^{\Lambda_c}(\nu,\nu')$ at the critical scale $\Lambda_c$. Hence, the frequency structure is not significantly changed in the symmetry-broken regime.
In the pairing channel there is even quantitative agreement between $\tilde V_p^{\Lambda_c}(\nu,\nu')$ and $\mathcal{A}_p^\Lambda(\nu,\nu')$. This indicates that $\tilde V_p^{\Lambda_c} \Pi_p^\Lambda$ becomes small for $\Lambda \to 0$, see Eq.~(\ref{eq:A_X}), which must be due to a strong suppression of $\Pi_p^\Lambda$ by the gap formation.

For large frequencies and away from the special lines $\nu' = \pm \nu$, the reduced magnetic interaction $\tilde V_m^{\Lambda_c}(\nu,\nu')$ tends to the bare Hubbard coupling $U$, while the reduced pairing interaction $\tilde V_p^{\Lambda_c}(\nu,\nu')$ decays to zero. The latter behavior is the reason for the decay of the pairing gap at large frequencies described above.


\subsection{Superfluid stiffness and phase diagram}

We finally compute the superfluid phase stiffness, which allows us to estimate the Kosterlitz-Thouless temperature $T_{\rm KT}$ for the onset of superconductivity. Together with the temperature $T^*$ for the onset of antiferromagnetism, we can thus draw a phase diagram in the plane spanned by doping and temperature.

\begin{figure}[t!]
 \includegraphics[scale=0.55]{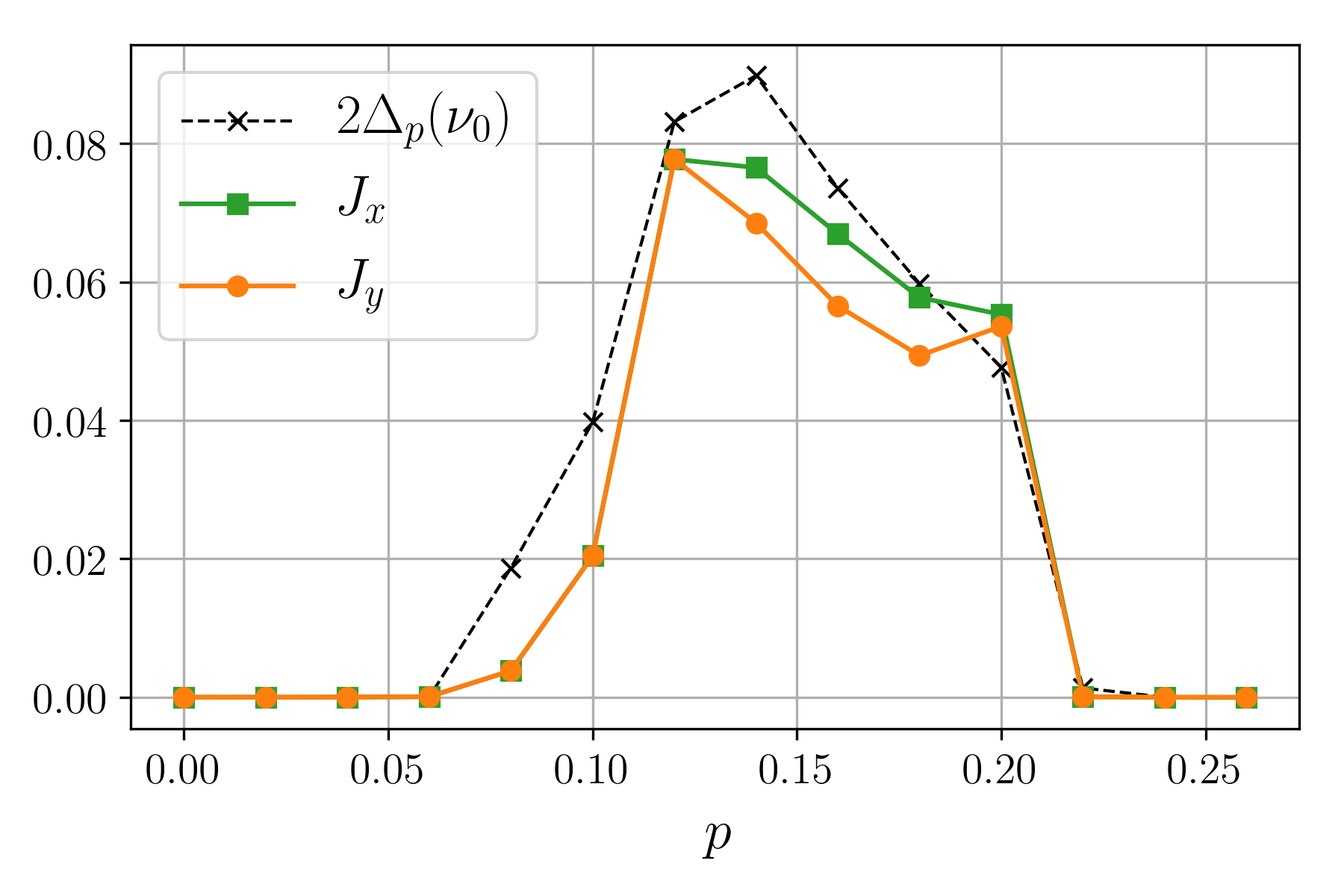}
 \caption{Phase stiffness in $x$- and $y$-direction as a function of doping at fixed
 temperature $T=0.027t$. The pairing gap amplitude $2\Delta_p(\nu_0)$ is also shown for
 comparison.}
 \label{fig:Stiffness}
\end{figure}
A general expression for the phase stiffness in a mean-field state with coexisting spin-singlet superconductivity and antiferromagnetism (N\'eel or spiral) has been derived in a recent work by Yamase and one of us. \cite{Metzner2019}
The phase stiffness is fully determined by the bare dispersion relation and the magnetic and pairing gaps. The gaps have been assumed to be frequency independent in the derivation. We therefore neglect the frequency dependence of the gaps, and insert the gap at the lowest Matsubara frequency $\nu_0 = \pi T$.
In a spiral state with an ordering vector $\bQ = (\pi - 2\pi\eta,\pi)$ with $\eta > 0$ the phase stiffness in $x$- and $y$-direction is slightly different.
In Fig.~\ref{fig:Stiffness} we plot the phase stiffnesses $J_x$ and $J_y$ as a function of doping at the fixed temperature $T = 0.027t$. The pairing gap amplitude from Fig.~\ref{fig:gaps} is also reproduced for direct comparison. In the N\'eel state the stiffness is isotropic, $J_x = J_y$, while in the spiral regime $J_y$ is slightly smaller than $J_x$.
The stiffness and the gap amplitude have a comparable size in the regime where both are larger than the temperature, but the stiffness decreases much faster at low doping, since it is more strongly suppressed by thermal excitations than the gap. By contrast, a static fRG calculation at zero temperature indicated that in the ground state the stiffness decreases slightly more slowly than the gap amplitude upon approaching half-filling from the hole-doped side. \cite{Metzner2019}
Note that, in general, there is no direct relation between the size of the gap and the size of the phase stiffness in a superconductor.

\begin{figure}[t!]
 \includegraphics[scale=0.55]{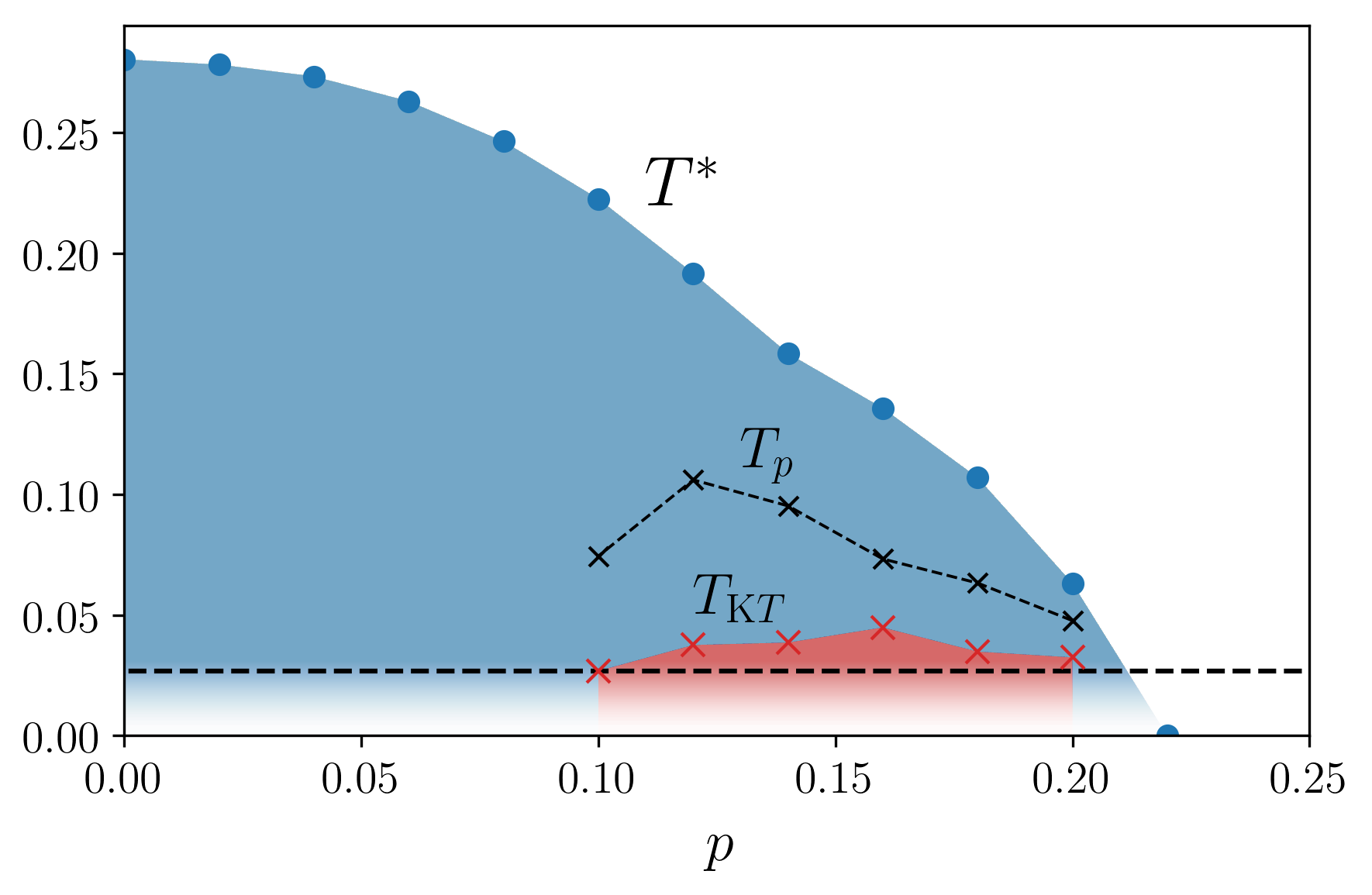}
 \caption{$(p,T)$ phase diagram with the critical temperature for the onset of
 antiferromagnetism $T^*$, the pairing temperature $T_p$, and the
 Kosterlitz-Thouless temperature $T_{\rm KT}$.
 The dashed horizontal line indicates the lowest temperature $T = 0.027t$ our
 present dynamical fRG code can access.}
 \label{fig:Phasedia}
\end{figure}
In a two-dimensional system, the thermal phase transition between the superfluid and the normal phase is a Kosterlitz-Thouless transition associated with topological excitations (vortices). \cite{Chaikin1995} Magnetic order or (non-critical) magnetic fluctuations do not affect the universal properties of this transition. In an isotropic system, the transition temperature $T_{\rm KT}$ is related to the phase stiffness $J$ by the universal relation
$T_{\rm KT} = \frac{\pi}{2} J(T_{\rm KT})$. \cite{Chaikin1995}
In the spiral state the phase stiffness is slightly anisotropic, that is, we have $J_x \neq J_y$, while there is still a unique transition temperature. Generalizing the relation between $T_{\rm KT}$ and $J$ to anisotropic systems by a simple rescaling of the length scales in the phase action, we find
\begin{equation} \label{eq:T_KT}
 T_{\rm KT} = \frac{\pi}{2} \sqrt{J_x(T_{\rm KT}) J_y(T_{\rm KT})} \, .
\end{equation}
Using this relation we are able to compute the Kosterlitz-Thouless temperature from the stiffnesses $J_\alpha(T)$, as long as $T_{\rm KT}$ is higher than the lowest temperature $T = 0.027t$ we can access. In Fig.~\ref{fig:Phasedia} we plot the resulting Kosterlitz-Thouless temperature as a function of doping, together with the critical temperature $T^*$ for the onset of antiferromagnetism. The latter is determined as the lowest temperature at which the fRG flow does not encounter any magnetic instability down to $\Lambda = 0$.
We also show the pairing temperature $T_p$ at which the pairing gap $\Delta_p$ vanishes. One can see that $T_p$ is much higher than $T_{\rm KT}$, especially at lower doping. Hence, a sizable temperature window with a pairing gap and superconducting fluctuations opens between $T_{\rm KT}$ and $T_p$. 

Our mean-field approximation yields magnetic long-range order for temperatures below $T^*$. General arguments and numerical studies show that fluctuations destroy this long-range order, giving rise to a pseudogap state with strong short-ranged magnetic correlations. Hence, $T^*$ should be interpreted as the onset temperature for pseudogap behavior. Implementing the fluctuations that turn the magnetically ordered state into a pseudogap state will be an interesting extension of our present theory.
Since we cannot access temperatures below $T = 0.027t$ with our present dynamical fRG code, we cannot calculate the Kosterlitz-Thouless temperature below 10 percent doping. Low temperature results from a static fRG flow indicate that $T_{\rm KT}$ vanishes linearly in doping upon approaching half-filling. \cite{Metzner2019}



\section{Conclusion}

We have performed a dynamical fRG analysis of magnetic order and superconductivity
in the two-dimensional repulsive Hubbard model at a moderate interaction strength $U=3t$.
A one-loop flow with coupled charge, magnetic and pairing interaction channels in the symmetric regime above the critical energy scale $\Lambda_c$ was combined with a mean-field approximation with decoupled reduced interactions in the symmetry-broken regime below $\Lambda_c$. The full frequency dependences of interaction vertices and gap functions were taken into account. The momentum dependences were approximated by suitable form factors. 

While magnetism appears as the leading instability at the critical scale $\Lambda_c$ in a broad doping range from half-filling to about 20 percent, robust pairing with a sizable pairing gap emerges between 10 and 20 percent doping, in coexistence with antiferromagnetism. The size of the pairing gap is slightly enhanced compared to results from a static fRG, probably as a consequence of the enhanced magnetic interactions.
The effective interactions exhibit strong frequency dependences, as already observed in previous dynamical fRG calculation, which were however limited to the symmetric regime. \cite{Husemann2012,Vilardi2017} The magnetic gap depends only weakly on frequency, while the pairing gap is peaked at the lowest Matsubara frequency and decays rapidly at higher frequencies.

We have also computed the superfluid phase stiffness and the Kosterlitz-Thouless transition temperature $T_{\rm KT}$ as a function of doping $p$.
Combining $T_{\rm KT}(p)$ with the onset temperature $T^*(p)$ for magnetism yields a phase diagram with a dome-shaped superconducting regime under the quasi-parabolic roof defined by $T^*(p)$. Including magnetic order parameter fluctuations would replace the ordered antiferromagnet in our theory by a pseudogap state with strong magnetic correlations at any finite temperature. The shape of $T^*(p)$ and $T_{\rm KT}(p)$ agrees qualitatively with the pseudogap temperature and the superconducting transition temperature, respectively, in high-$T_c$ cuprates.

Our work can be naturally extended in two directions. First, the fRG + MFT approach can be extended to the strongly interacting regime by using the dynamical mean-field solution \cite{Metzner1989,Georges1996} of the Hubbard model as a starting point for the fRG flow. The combination of dynamical mean-field theory and fRG was proposed some time ago, \cite{Taranto2014} and recently applied to the two-dimensional Hubbard model at strong coupling, albeit only in the symmetric regime up to the point where effective interactions diverge. \cite{Vilardi2019} Magnetic and pairing gap functions could now be computed by continuing the flow with reduced but dynamical interactions into the symmetry broken regime.

Second, the mean-field solution for magnetic order in the symmetry-broken regime could be improved by implementing thermal and quantum fluctuations of the spin orientation. A most promising route to do this is to treat the magnetic regime with an SU(2) gauge theory as recently developed by Scheurer et al. \cite{Scheurer2018} The critical temperature $T^*$ for magnetism obtained from the fRG flow assumes the role of the pseudogap temperature in that improved theory.


\section*{Acknowledgements}

We are very grateful to Thomas Sch\"afer and Hiroyuki Yamase for valuable discussions.


%
%
\end{document}